# Dependence of the Energy Transfer to Graphene on the Excitation Energy

Sebastian Mackowski, Izabela Kamińska

Institute of Physics, Faculty of Physics, Astronomy and Informatics, Nicolaus Copernicus University, Grudziadzka 5, 87-100 Torun, Poland

E-mail: mackowski@fizyka.umk.pl

**Abstract**

Fluorescence studies of natural photosynthetic complexes on a graphene layer demonstrate pronounced influence of the excitation wavelength on the energy transfer efficiency to graphene. Ultraviolet light yields much faster decay of fluorescence, with average efficiencies of the energy transfer equal to 87% and 65% for excitation at 405 nm and 640 nm, respectively. This implies that focused light changes locally the properties of graphene affecting the energy transfer dynamics, in an analogous way as in the case of metallic nanostructures. Demonstrating optical control of the energy transfer is important for exploiting unique properties of graphene in photonic and sensing architectures.

Energy transfer is one of the most fundamental processes at the nanoscale [1,2]. Whenever a donor is placed sufficiently close to an acceptor, assuming their spectral properties and relative orientations do not inhibit it, they can couple via electrostatic interactions and the energy is funneled down to the acceptor [1]. Such a scheme evolved in natural photosynthesis [3] for efficient capturing and transport of the sunlight energy, and has been recently implemented in artificial light-harvesting assemblies [4,5]. The energy transfer between molecules with precisely designed optical spectra has also been useful in studying and understanding molecular mechanisms responsible for protein folding [6], intracellular transport [7], etc., as the efficiency of this process is extremely sensitive to the distance between a donor and an acceptor [1].

The key spectral signature of the energy transfer is a decrease of the emission intensity of a donor at the expense of an acceptor with simultaneous shortening of a donor fluorescence decay time [1]. In fact, from the reduction of the decay constant it directly related to the efficiency of the energy transfer, which in molecular assemblies is independent of the donor excitation wavelength, as the rates of the energy transfer are typically a few orders of magnitude slower than intra-molecular transitions [1].

As an example, the energy transfer takes place between emitters and metallic surfaces or nanostructures [8,9,10], which feature strong plasmonic oscillations of free electrons. However, as metals typically fluoresce very weakly, the energy transferred from a dipole placed in their vicinity is dissipated non-radiatively, mainly by heat. In such cases the interaction between an emitter and a metallic nanostructure can be probed by monitoring decrease of donor emission intensity and shortening of its fluorescence decay. The resonant character of plasmon excitations in metallic nanostructures implies strong wavelength dependence of both metal-enhanced fluorescence and fluorescence quenching [9,11,12].

The uniqueness of graphene, a two-dimensional $sp^2$-hybridized carbon hexagonal lattice, has been advocated worldwide in the last decade, but predominantly in regard to its thermal, electrical, and mechanical properties [13]. Only recently the optical properties of graphene have emerged as highly attractive from the point of view of potential applications in photonics and optoelectronics [14]. One of the remarkable facets of graphene is its uniform absorption, which extends over the whole visible range down to the infrared [15]. With the absence of fluorescence, it renders graphene as an exceptional acceptor in devices that utilize energy and/or electron transfer.

Recently, energy transfer in graphene-based assemblies has been studied both experimentally and theoretically [16,17]. The two-dimensional character of graphene results in a weaker distance dependence of the energy transfer efficiency ($\sim d^{-4}$) as compared to a three-dimensional case ($\sim d^{-6}$) [16]. This difference has been verified experimentally by studying rhodamine dyes on graphene [18]. The results indicate that this coupling can be exceptionally strong, with the decay rates being enhanced by up to 90 times.

In this work we aim at demonstrating that the efficiency of the energy transfer to graphene depends on the excitation wavelength. To this end a molecular system which features broad absorption, is a few-nm large, and fluoresces is required. Some of the naturally evolved photosynthetic complexes fulfill these conditions. For photosynthetic complexes deposited directly onto a graphene layer we observe decrease of the fluorescence intensity accompanied with reduction of the decay time. The key finding that emerges from these experiments is a strong dependence of the efficiency of the energy transfer on the excitation wavelength: the effect is much stronger for 405 nm than for 640 nm excitation. We conclude that energy quenching in graphene is driven not only by dipole-dipole interaction, but also a mechanism associated with light-induced oscillations of electrons in graphene. Indeed, exciting electrons in graphene has an effect of its dissipative efficiency, which opens avenues in interfacing electronic and plasmonic character of graphene in hybrid nanostructures and control the electronic dynamics of such systems with light.

In order to probe the interactions with graphene, we use peridinin-chlorophyll-protein (PCP) complex from algae *Amphidinium carterae*. It is a water-soluble light-harvesting complex, whose structure was determined with 1.7A resolution [19]. Optical spectra of PCP in solution are displayed in Fig. 1. The absorption spans over the whole visible range with the main band (from 350 to 550 nm) attributed mainly to peridinin absorption. Chlorophylls absorb through the $Q_y$ band around 670 nm and the Soret band at 437 nm [20]. The emission of PCP complex is associated with fluorescence of chlorophylls and occurs at 673 nm, with a 30% quantum yield and a decay time constant of 4 ns [21]. The decay time constant oft he fluorescence emisison is around 3 orders of magnitude shorter than any oft he intra-complex transfer times, within 3 ps, regardless of the excitation wavelength, the energy is transfered to the Qy band of chlorophyll molecules [20].

Graphene substrates were purchased from Graphene Supermarket®. We used 1cm × 1cm p-doped silicon wafers with a single layer graphene deposited using Chemical Vapor Deposition on a 285 nm thick silicon dioxide layer. The presence of graphene monolayer on the substrates was confirmed with Raman spectrum measurement.

For optical experiments we used highly diluted (optical density of 0.009 at 671 nm, concentration less that 10 µM) aqueous solution of PCP complexes. Such a low concentration is very important as on the one hand it strongly reduces inner filter effect, but this also yields a thin layer of PCP complexes on a graphene surface. As a result, we minimize the fraction of PCP that is not coupled to graphene, thus takes no part in the energy transfer.

Absorption and fluorescence spectra of PCP solution were obtained at room temperature using Cary 50 UV-Vis spectrophotometer (Varian) and Fluorolog 3 spectrofluorimeter (Jobin-Yvon). Spectrally- and time-resolved fluorescence measurements were performed using a home-built confocal fluorescence microscope described in detail in [22]. The sample was placed on a piezoelectric translation stage. We used pulsed laser excitation at 405 nm and 640 nm (repetition rate 20 MHz, average power 30 µW, power density ~1MW/cm$^{-2}$). Importantly, PCP can be efficiently excited at 405 nm (Soret band), and 640 nm (excited states of chlorophylls). The laser beam was focused on the sample by LMPlan 50x objective (Olympus) with a numerical aperture 0.5. Fluorescence was first filtered by a longpass filter (HQ665LP Chroma) and then the spectra were detected with Andor iDus DV 420A-BV CCD camera coupled to an Amici prism. Time-resolved measurements were performed by time-correlated single photon counting technique using an SPC-150 module (Becker & Hickl) with fast avalanche photodiode (idQuantique id100-20) as a detector. In order to select appropriate wavelength range, we used an additional bandpass filter (FB670/10 Thorlabs).

When compared with the reference, fluorescence spectra for PCP on graphene (as shown in Fig. S1) indicate strong reduction of the emission intensity, an effect we can tentatively attribute to the energy transfer [23]. At the same time, the shape of the PCP emission spectrum on graphene remains unaffected, and is identical to previously published [21], which indicates that the protein is intact and that all the energy transfer pathways are active in the photosynthetic complex upon deposition on graphene. This observation implies graphene suitability for interfacing with functional biological systems.

The assumption that the observed reduction of the emission intensity of PCP complexes deposited on graphene is due to the energy transfer from the chlorophylls in the photosynthetic complex to graphene is supported by a strong decrease of the fluorescence decay, as shown in Fig. 2a and 2b for 405 nm and 640 nm excitations, respectively. Three transients shown for each excitation wavelength were measured at different locations across the sample. For both excitation wavelengths fluorescence decays much faster than for the reference. In addition, the transients measured for PCP complexes on graphene, feature high inhomogeneity: there are decays that are relatively long, as well as comparable with the temporal resolution of the optical setup. While the results shown in Fig. 3a and 3b display transients close to average as well as borderline cases, we note that for the 405 nm excitation the majority of measured decays is very fast, while exciting with 640 nm yields considerably longer decays.

The distribution of fluorescence transients indicates that PCP complexes probed in every measurement couple slightly different to the graphene layer. It is expected as we have not incorporated any control of the separation between PCP complexes and graphene to make the interaction more uniform across the substrate. Furthermore, it has been shown that for graphene deposited on silica, the local structure of graphene is also quite inhomogeneous with islands of high and low mobility of carriers [24]. We might therefore assume that such a non-uniformity contributes also to the observed spreading of fluorescence transients, although the scale of these inhomogeneities is less than 100 nm, which is significantly less than spatial resolution of the fluorescence microscope. Regardless, it is striking that most of them exhibit almost monoexponential behavior. It could indicate that majority of PCP complexes with the laser spot is coupled to graphene with a comparable strength, although we clearly see distribution of lifetimes indicating the presence of sub-ensembles in our structure [1]. Each of the lifetimes would then be attributed to the PCP complexes coupled to graphene with different efficiency leading also to a different energy transfer rate. Certainly we do not observe long (~4 ns) component attributable to PCP complexes not coupled to graphene.

We can conclude that the fluorescence experiments carried out for PCP deposited on a graphene surface demonstrate shortening of the fluorescence decay, which is accompanied with a decrease of the overall fluorescence intensity. Taken together, these observations strongly suggest that the energy absorbed by PCP complexes, both at 405 nm and 640 nm, after being transferred between and within the pigments comprising the light-harvesting complex, is efficiently dissipated into the graphene layer. In general such effects can also be attributed to electron transfer from photosynthetic complex to graphene, as observed recently in [25]. However in this case a charge-separating complex, that is Photosystem I, was used and the whole structure was immersed in properly chosen electrolyte in order to facilitate the electron transfer. As the PCP complex is just a light-harvesting antenna, we can exclude such a scenario in our structure.

As pointed out earlier, in molecular systems, where the interaction leading to the energy transfer takes place between two dipole moments, the decay constant of a donor in the presence of an acceptor is independent of the excitation wavelength used for exciting the donor. This is a reminiscence of the fact that light has no effect on the surrounding of the molecules participating in the energy transfer. In contrast, in the case of PCP complexes deposited on graphene, the fluorescence decay strongly depends on the excitation wavelength, as displayed in Fig. 4, where histograms of decay times measured for 405 nm and 640 nm excitations are compared. The average decay times are equal to 0.5 ns and 1.4 ns, respectively, what results in corresponding energy transfer efficiencies of 87% and 65%. Such a clear influence of the excitation wavelength on the energy transfer indicates that in addition to populating PCP excited states (the pigments embedded within the protein), light changes also the local surroundings, presumably the properties of graphene. In a way, although the analogy is certainly not complete, strong dependence of the donor decay on the excitation wavelength, is similar to metallic systems, where in order to facilitate the energy transfer it is necessary to excite particular resonance, i.e. plasmon resonance.

On a final note, by integrating fluorescence transients we can estimate total emission intensity of PCP complexes, and therefore, obtain correlation between fluorescence decay and intensity. The result plotted in Fig. 5 shows that for PCP complexes deposited on graphene, excitation with both 405 nm and 640 nm results in strongly correlated values of fluorescence decay times and integrated intensities. Namely, shorter decay times are correlated with lower intensities. Correlation coefficients (Pearson's coefficients) estimated for both excitation wavelengths are approximately 0.7. As both these quantities are spectral signatures of the energy transfer, correlation between them strongly suggests that decrease of emission intensity is indeed directly attributed to the energy transfer from PCP to graphene.

Although elaboration of the processes responsible for the observed excitation energy dependence of the energy transfer to graphene requires further experiments, in analogy to plasmonic nanostructures, we can propose a scenario that in principle can explain this effect. Electrons excited in graphene by light can oscillate, similarly as in a metallic nanoparticle, in a confined space defined by a monolayer of graphene on the one hand, and the size of the laser spot, on the other. As the latter is around 1 micron in a diameter, it is comparable to the wavelength of light used in the experiment. Therefore, locally excited graphene layer can be considered analogous to a metallic nanostructure with strong sensitivity of its dissipative properties on the excitation wavelength. Small variation of the excitation wavelength (less than 100 nm, characteristic of most organic dyes) is not sufficient to demonstrate this unique effect in a convincing way, therefore an emitter that can be excited across the wide range of wavelengths is required.

In conclusion, we observe strong dependence of the energy transfer efficiency in hybrid assemblies composed of natural photosynthetic complexes and graphene on the excitation wavelength. High energy excitation (405 nm) results in energy transfer efficiency of 87%, while low energy one (640 nm) yields efficiency of 65%. This proof-of-concept result, in addition to paving a way towards designing novel graphene-based structures for photosynthetic energy harvesting and conversion, indicates that the energy transfer in a hybrid assembly can be controlled by light.

We thank Prof. Eckhard Hofmann (Bochum University, Germany) for providing PCP complexes. Research was supported by the project number DEC-2013/10/E/ST3/00034 from the National Science Center (NCN) and the grant ORGANOMET No: PBS2/A5/40/2014 from the National Research and Development Center (NCBiR).

**Fig. 1.** Optical spectra of PCP complexes in solution: red and black curve corresponds to absorption and emission, respectively.

**Fig. 2.** Examples of normalized fluorescence transients measured for PCP complexes deposited on a single layer graphene measured for excitation (a) 405 nm and (b) 640 nm. Green lines correspond to the reference.

**Fig. 3.** (a) Histograms of the decay times obtained for PCP on single layer graphene for excitation 405 nm (blue) and 640 nm (red). (b) Correlation between fluorescence intensity and the decay time measured for PCP on graphene for excitation 405 nm (blue) and 640 nm (red).

Fig. 1.

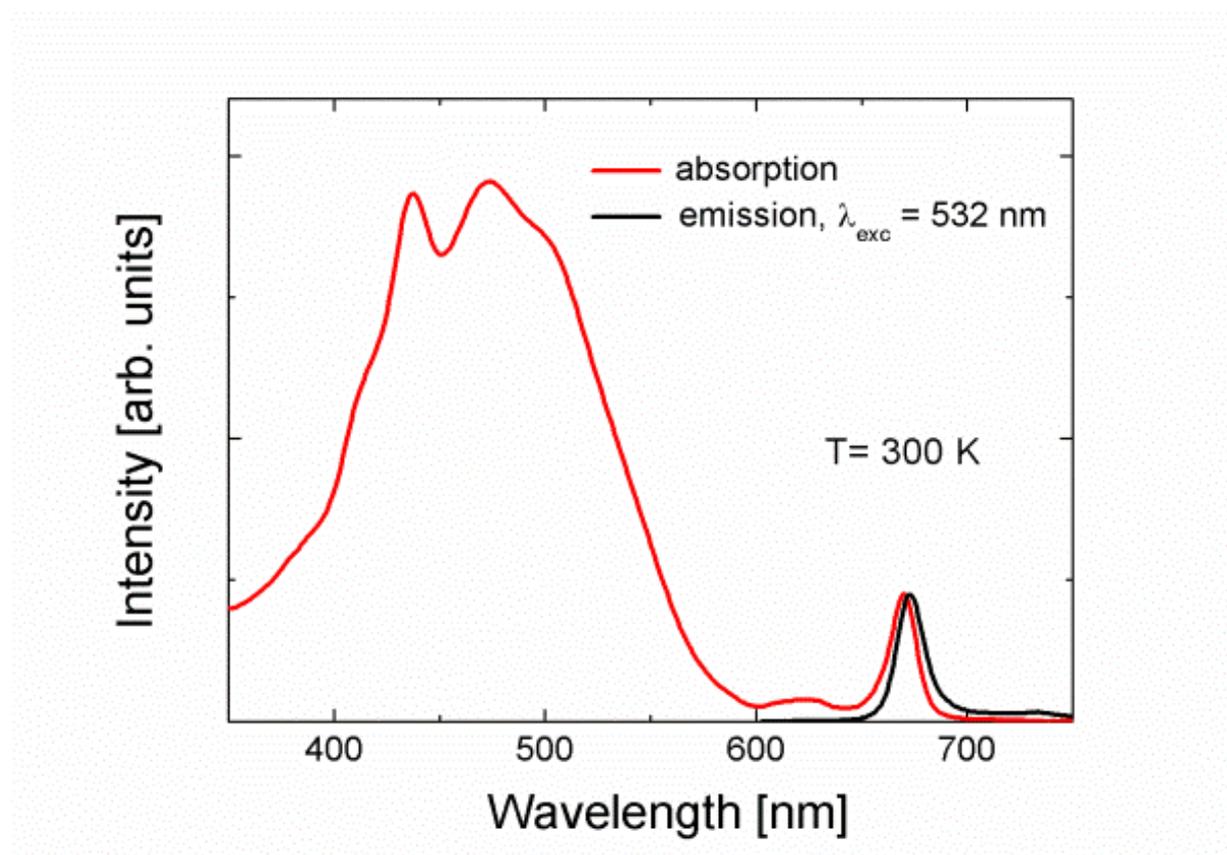

Fig. 2.

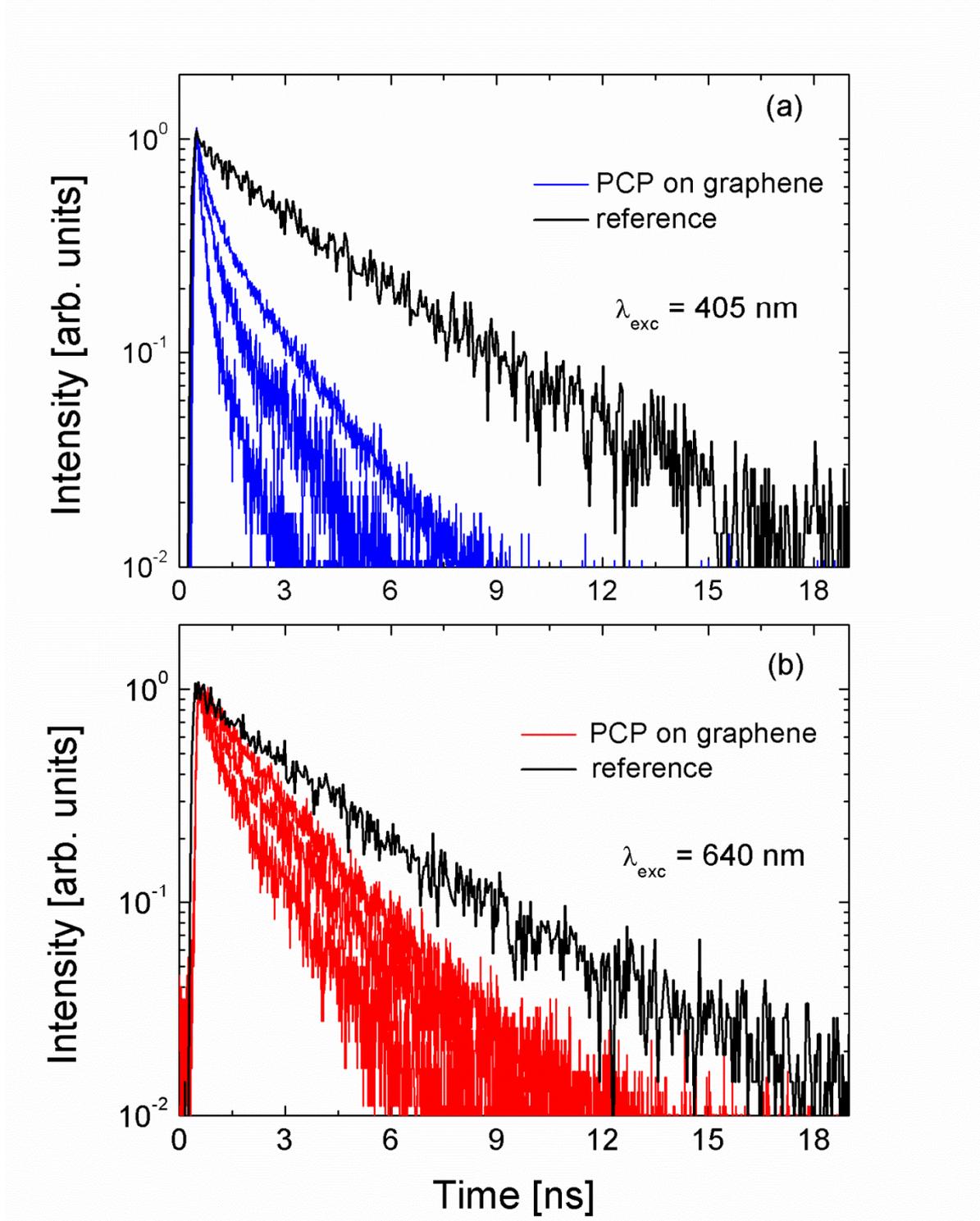

Fig. 3.

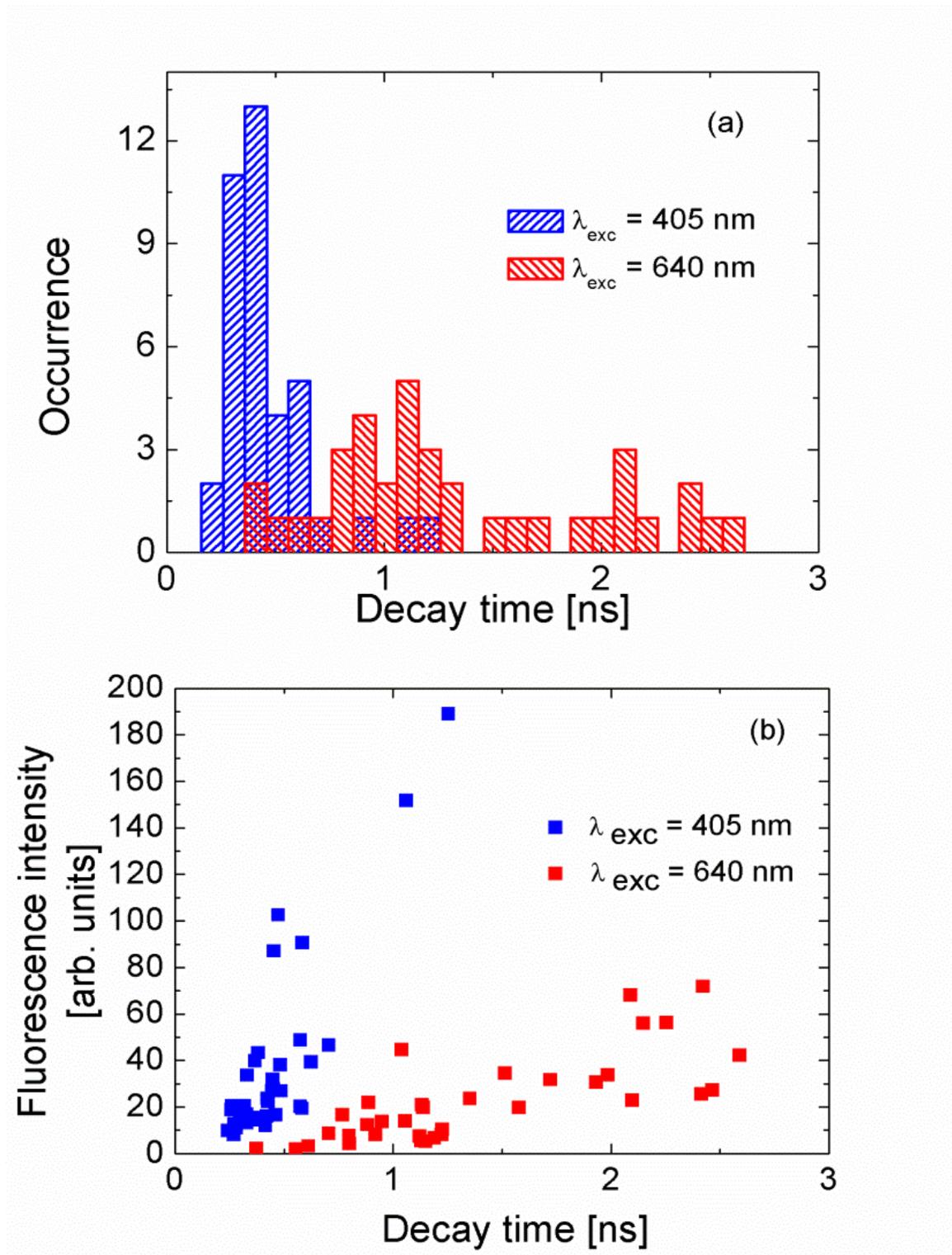